\documentstyle[sprocl,amsmath,amssymb]{article}

\input{psfig}

\def\be{\begin{equation}}
\def\ee{\end{equation}}
\def\bea{\begin{eqnarray}}
\def\eea{\end{eqnarray}}

\newcommand{\del}{\partial}
\newcommand{\oa}[1]{${\cal O}(a^{#1})$}
\newcommand{\csw}{c_{\text{sw}}}
\newcommand{\order}{{\mathcal O}}

\newcommand{\beq}{\begin{equation}}
\newcommand{\eeq}{\end{equation}}
\newcommand{\beqa}{\begin{eqnarray}}
\newcommand{\eeqa}{\end{eqnarray}}

\newcommand{\pslash}{\not\!p}
\newcommand{\kslash}{\not\!k}

\newcommand{\qhat}{\hat{q}}

\newcommand{\ebox}[2]{\epsfxsize=#1 \epsfbox[10 30 560 590]{#2}}

\begin{document}

\title{GLUONS, QUARKS, AND THE TRANSITION FROM NONPERTURBATIVE
              TO PERTURBATIVE QCD
          }

\author{\underline{ANTHONY G.\ WILLIAMS}, FREDERIC D.R.\ BONNET,
          PATRICK O.\ BOWMAN,
          DEREK B.\ LEINWEBER, JON IVAR SKULLERUD, AND JAMES M.\ ZANOTTI}

\address{CSSM and Department of Physics and Mathematical
             Physics, University of Adelaide, Australia 5005\\
             E-mail: {\tt awilliam@physics.adelaide.edu.au}  }


\maketitle\abstracts{
Lattice-based investigations of two
fundamental QCD quantities are described, namely
the gluon and quark propagators in Landau gauge.
We have studied the Landau gauge gluon
propagator using a variety of lattices with spacings from $a = 0.17$
to 0.41 fm to explore finite volume and discretization
effects.  We also introduce the general method of ``tree-level correction''
to minimize the effect of lattice artefacts at large momenta.
We demonstrate that it is possible to obtain scaling behavior over a
very wide range of momenta and lattice spacings and
to explore the infinite volume and continuum limits of the Landau-gauge
gluon propagator.  These results confirm
the earlier conclusion that the Landau gauge gluon propagator is
infrared finite.
We study the Landau gauge quark propagator in quenched QCD using two
forms of the $\order (a)$-improved propagator with the
Sheikholeslami-Wohlert quark action with the nonperturbative
value for the clover coefficient $\csw$ and mean-field 
improvement coefficients in our improved quark propagators.  
We again implement an appropriate form of tree-level correction.
We find good agreement between our improved quark propagators.
The extracted value of infrared quark
mass in the chiral limit is found to be $300\pm 30$ MeV.
We conclude that the momentum regime where
the transition from nonperturbative to perturbative QCD occurs
is $Q^2\simeq 4$~GeV$^2$.}

\section{Introduction}
\label{sec:intro}

Lattice gauge theory is currently the only known ``first principles''
approach to studying nonperturbative QCD.  
It is therefore important for lattice
QCD to provide constraints and guidance for the construction
of quark-based models\cite{DSE_review}
and to provide an indication of the momentum
regime at which we can expect perturbative QCD to become applicable.
The quark and gluon propagators are two of the most fundamental
quantities in QCD. There has been considerable interest in the
infrared behavior of the 
gluon propagator as a probe into the mechanism of confinement and by studying 
the scalar part of the quark propagator, the mass function, we can gain 
insight into the mechanisms of chiral symmetry breaking. 
Both are used as input for other quark-model
calculations.

\section{Gluon Propagator}

We use an ${\cal O}(a^2)$ tree-level, tadpole-improved action\cite{Weisz83}
and for the tadpole (mean-field)
improvement parameter we use the plaquette measure\cite{tadpole}.
A full description and discussion of the gluon propagator results
summarized here can be found
elsewhere.\cite{LandauGaugeDE,long_glu,big_vol_glu}

\begin{table}[tb]
\centering
\begin{tabular}{cccccc}
   & Dimensions	     & $\beta$ & $a$ (fm) & Volume $\text{(fm}^4\text{)}$ &
Configurations \\
\hline
1w & $16^3\times 32$ &   5.70  &   0.179  & $2.87^3 \times 5.73$  & 100 \\
1i & $16^3\times 32$ &   4.38  &   0.166  & $2.64^3 \times 5.28$  & 100 \\
2  & $10^3\times 20$ &   3.92  &   0.353  & $3.53^3 \times 7.06$  & 100 \\
3  & $8^3 \times 16$ &   3.75  &   0.413  & $3.30^3 \times 6.60$  & 100 \\
4  & $16^3\times 32$ &   3.92  &   0.353  & $5.65^3 \times 11.30$ & 100 \\
5  & $12^3\times 24$ &   4.10  &   0.270  & $3.24^3 \times 6.48$  & 100 \\
6  & $32^3\times 64$ &   6.00  &   0.099  & $3.18^3 \times 6.34$  &  75 \\
\end{tabular}
\caption{Details of the lattices used to calculate the gluon propagator.  
Lattices 1w and 1i have the same dimensions and approximately the same lattice
spacing, but were generated with the Wilson and improved actions respectively.
Lattice 6 was generated with the Wilson action.}
\label{table:latlist}
\end{table}

\begin{figure}[ht]
\begin{center}
\epsfig{figure=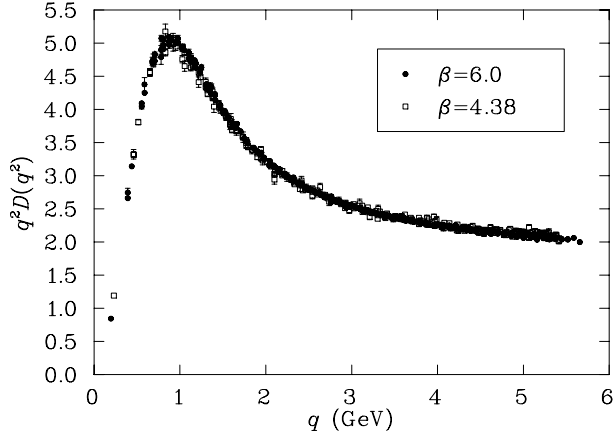,angle=90,width=8cm}
\end{center}
\caption{Comparison of the gluon propagator from the finest improved
lattice (lattice 1i, $\beta = 4.38$) and the finest Wilson lattice 
(lattice 6, $\beta=6.0$).  Data has been cylinder cut and the appropriate
tree-level corrections have been applied.  The data from lattice 6 is
half-cut whereas lattice 1i displays the full Brillouin zone.  We have
determined $Z_3(\text{improved}) / Z_3(\text{Wilson}) = 1.08$ by
matching the vertical scales of the data.}  
\label{fig:Comp1i_6}
\end{figure}

\begin{figure}[ht]
\begin{center}
\epsfig{figure=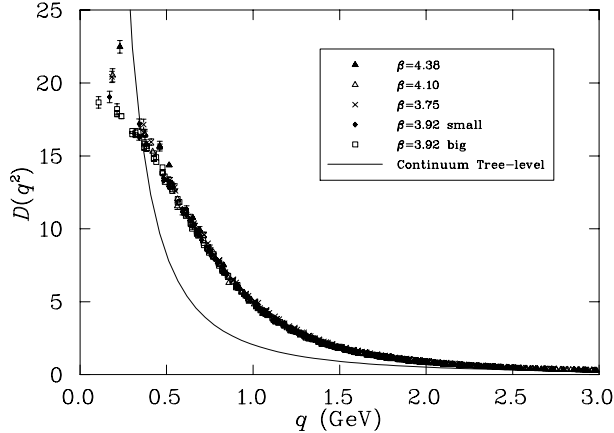,angle=90,width=8cm}
\end{center}
\caption{Comparison of the gluon propagator generated with an improved action
on five different lattices.  We find good agreement down to 
$q \simeq 500$ MeV.  At the lowest accessible momenta the data points drop
monotonically with increasing volume, but the lowest point (on the largest 
lattice) shows signs of having converged to its infinite volume value.
For comparison with perturbation theory, a plot of the continuum, tree-level 
gluon propagator (i.e., $1 / q^2$ appropriately scaled) has been included.}
\label{fig:AllProps}
\end{figure}

\begin{figure}[ht]
\begin{center}
\epsfig{figure=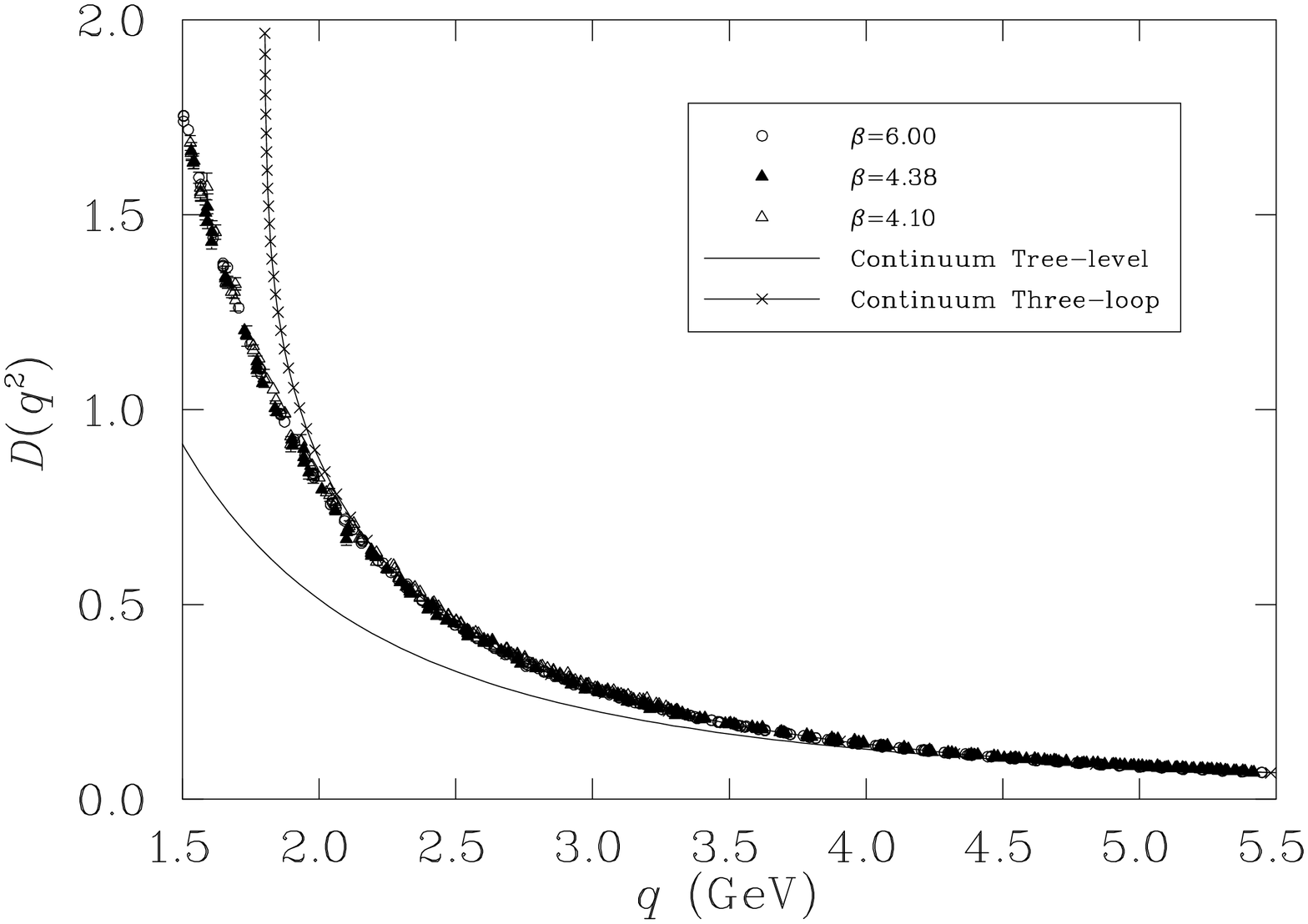,angle=90,width=8cm}
\end{center}
\caption{Comparison of the lattice gluon propagator with that obtained from
perturbation theory, in the ultraviolet to intermediate regime.  The continuum
expressions are tree-level (i.e., $1 / q^2$ appropriately scaled) and
the three-loop perturbative QCD expression.}
\label{fig:pert_vs_latt}
\end{figure}

Gauge fixing on the lattice is achieved by maximizing a functional, the 
extremum of which implies the gauge fixing condition.  The usual Landau 
gauge fixing functional implies that $\sum_\mu \del_\mu A_\mu
= 0$ up to \oa{2}.  To ensure that gauge dependent quantities are also
\oa{2} improved, we implement the analogous \oa{2} improved gauge
fixing.\cite{LandauGaugeDE}
The dimensionless lattice gluon field $A_{\mu}(x)$ is calculated from
the link variables in the usual way, 
which agrees with the continuum to ${\cal O}(a^2)$.
We then calculate the scalar part of the propagator
\begin{equation}
D(x-y) = \sum_\mu \langle A_\mu(y) A_\mu(x) \rangle \, .
\end{equation}
To isolate the nonperturbative behavior of the gluon propagator, we
can divide the propagator by its lattice tree level form (i.e., that
of lattice perturbation theory).\cite{long_glu}
For the momentum space gluon propagator $D(q^2)$, we see
that in the continuum $q^2D(q^2)$ will approach a constant up to 
logarithmic corrections as $q^2\to\infty$ because of asymptotic
freedom.  The continuum tree-level propagator is $1/q^2$.
We also expect asymptotic freedom on the lattice despite
finite lattice spacing artefacts.  We {\em define} the lattice
$q_\mu$ such that the lattice $D^{\rm tree}(q)\equiv 1/q^2$,
and use this momentum throughout. This is referred to as tree-level
correction and we have seen that it significantly reduces discretization
arrors at large momenta.  For the two actions considered here,
this means that we work with the momentum variables defined as
\begin{equation}
q_{\mu}^W \equiv \frac{2}{a} \sin\frac{\qhat_{\mu} a}{2},
\hskip1cm
q_\mu^I \equiv \frac{2}{a}\sqrt{ \sin^2 
	\Bigl( \frac{\qhat_\mu a}{2} \Bigr)
	+ \frac{1}{3}\sin^4 \Bigl( \frac{\qhat_\mu a}{2} \Bigr) 
	} \, ,
\label{eq:latt_momenta}
\end{equation}
for the Wilson and improved actions respectively.
All figures (quark and gluon propagators) have a cylinder cut imposed upon 
them, i.e.\ all momenta must lie close to the lattice diagonal.
In Table~\ref{table:latlist} we show the various lattices that we have
studied for the gluon propagator.  In Fig.~\ref{fig:Comp1i_6} 
we plot $q^2 D(q^2)$ for a fine unimproved Wilson action and for
our finest improved action.  Despite having very different lattice
spacings the agreement is excellent for the entire intermediate
and high-momentum regime.  The small discrepancy in the deep infrared
due to finite volume effects is not apparent in this way of plotting
that data.  We plot $D(q^2)$ for five different lattice in
Fig.~\ref{fig:AllProps} and see pleasing agreement for the results.
Note that we are plotting bare quantities only and there is thus an
overall wavefunction renormalization for the gluon propagator
(i.e., $Z_3(\mu,a)$ for the renomalization point $\mu$).  The vertical
scale is thus unimportant and only the variation with momentum
is relevant.
This way of presenting the data shows that there is a small residual
finite volume dependence, where the infrared gluon propagator is
{\em decreasing} with increasing lattice volume $V$.  We have performed
a fit as a function of $1/V$ and have seen that the large volume
$\beta=3.92$ lattice gives results which are very close to the
infinite volume limit.
In Fig.~\ref{fig:pert_vs_latt} we plot $D(q^2)$ in the intermediate
and ultraviolet regime and compare with the three-loop perturbative
QCD form.  We see that the above $q\sim 2$~GeV the agreement is
excellent, but that below this momentum scale nonperturbative effects
are becoming apparent.

\section{Quark Propagator}

The Landau gauge quark propagator results summarized here have been
presented and discussed in more detail
elsewhere.\cite{quark_prop}
All ${\cal O}(a)$ errors in the fermion action can be removed by adding
appropriate terms to the Lagrangian\cite{Luscher:1996sc,Dawson:1997gp}.
It is then usual to perform appropriate field
transformations to improve the quark operators
as well.\cite{Heatlie:1991kg}


\begin{figure}[ht]
\begin{center}
\epsfig{figure=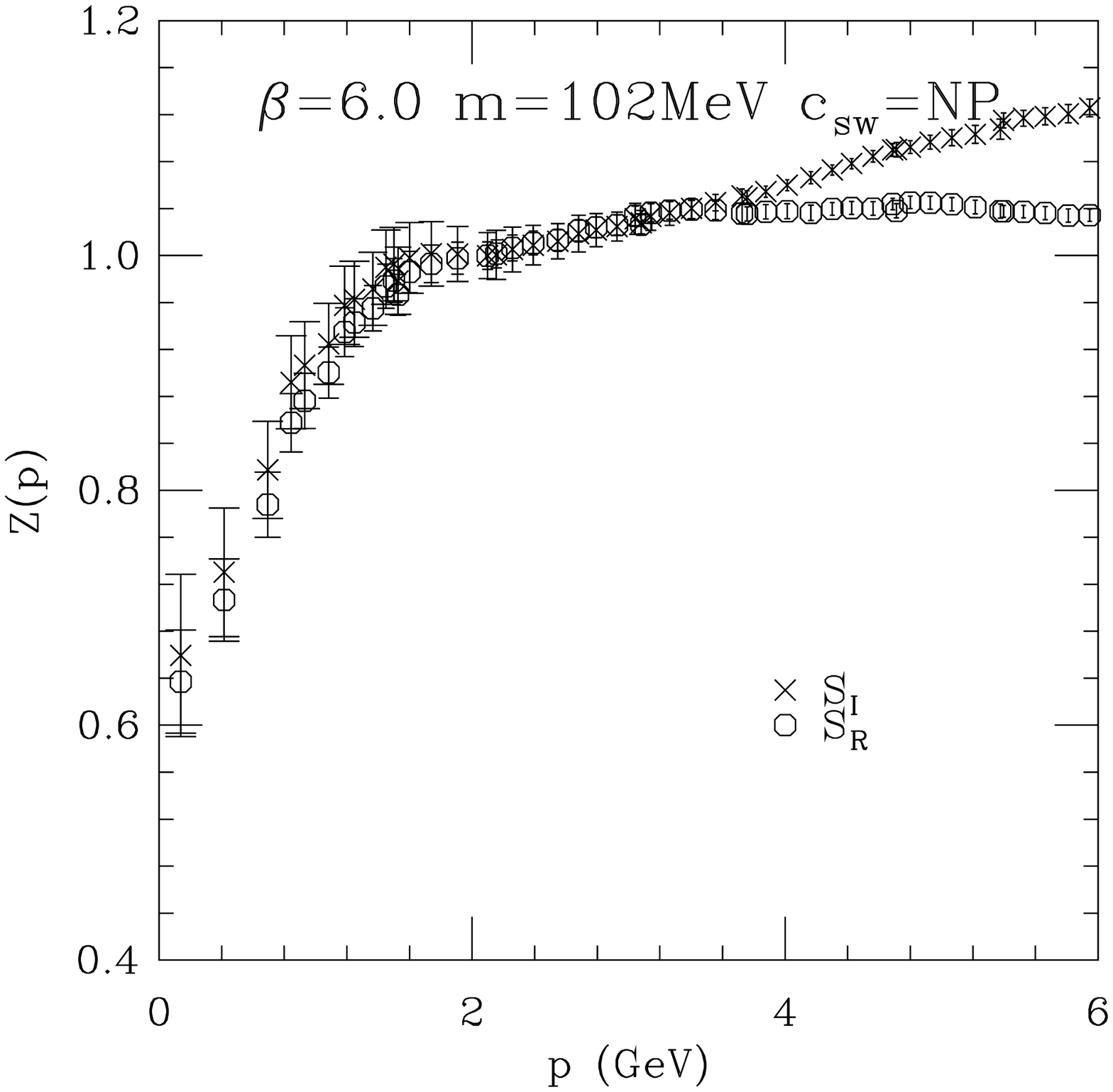,angle=0,width=7cm}
\end{center}
\caption{The tree-level corrected $Z(p)$ for $\csw=\text{NP}$ and
$\kappa=0.137$.
The figure shows $Z(p)$ vs.\ momentum in
physical units, and after rescaling (``renormalizing'') so that
$Z(2.1 \text{GeV})=1$. 
The infrared agreement after rescaling is very good and we take the tree-level
corrected $Z(p)$ from $S_R$ with the nonperturbative $\csw$ as the best
estimate for this quantity.
}
\label{fig:z_np_compare}
\end{figure}

\begin{figure}[ht]
\begin{center}
\setlength{\unitlength}{0.85cm}
\setlength{\fboxsep}{0cm}
\begin{picture}(14,7)
\put(0,0){\begin{picture}(7,7)\put(-0.9,-0.4){\ebox{6cm}{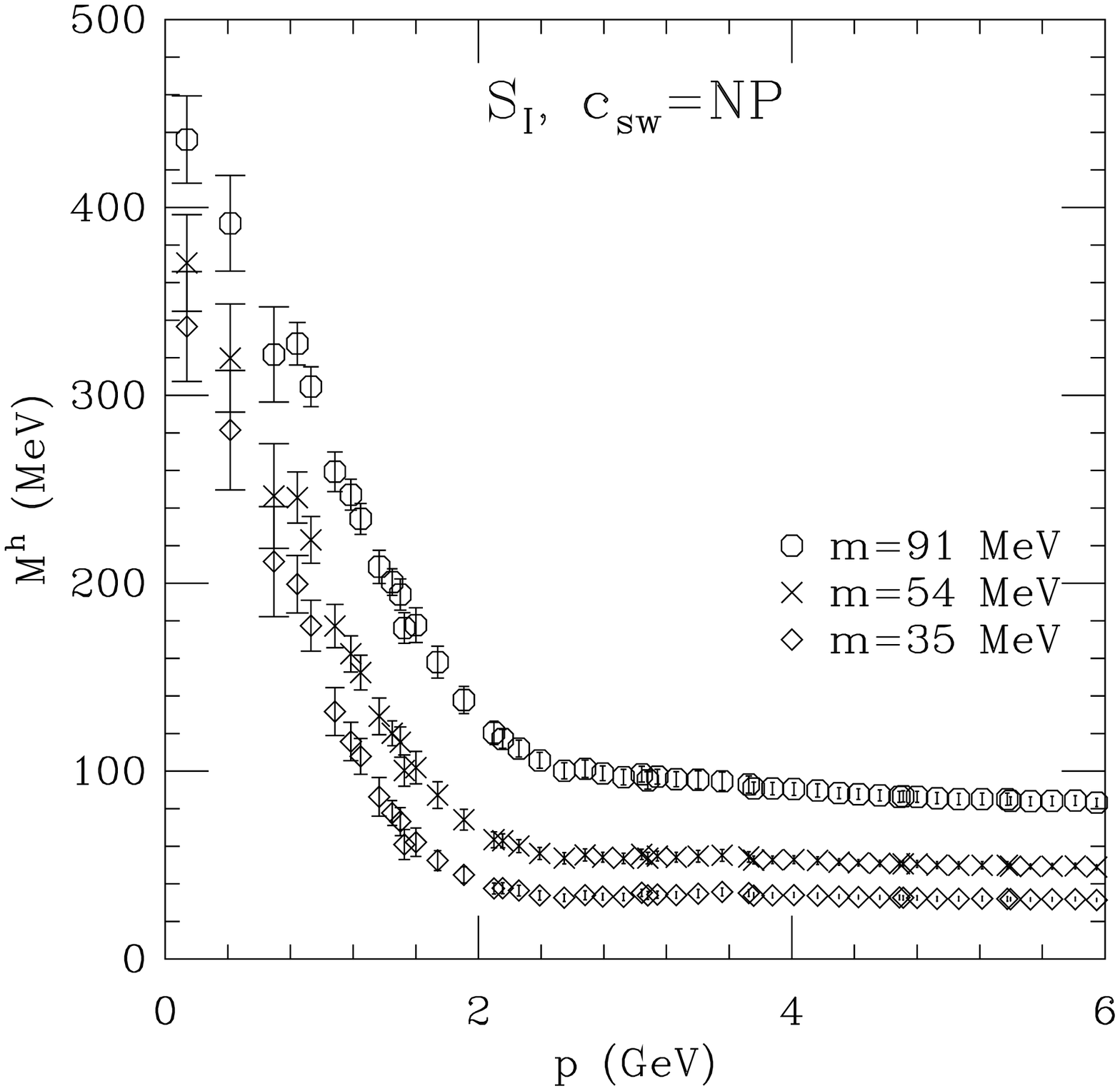}}\end{picture}}
\put(7,0){\begin{picture}(7,7)\put(-0.9,-0.4){\ebox{6cm}{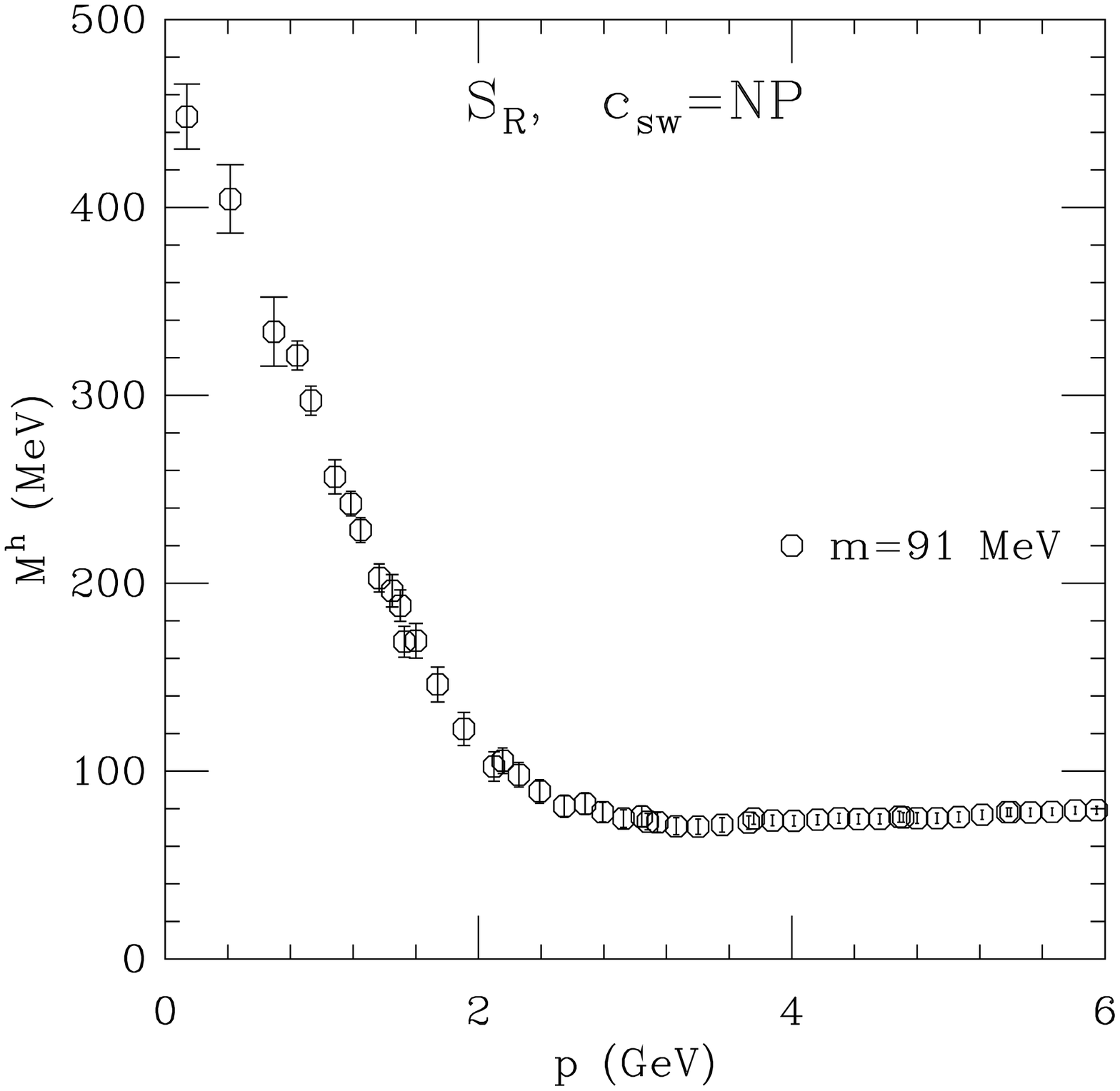}}\end{picture}}
\end{picture}
\end{center}
\caption{
The tree-level corrected $M(p)$, for $\csw=\text{NP}$, using
$S_I$ (left) and $S_R$ (right).  We find good
agreement between the two data sets, both in the infrared and
ultraviolet.  The residual disagreement at intermediate momenta is a
pointer to lattice artifacts that we have not brought under full
control, even with nonperturbative improvement and tree-level
correction. 
}
\label{fig:m_np_phys}
\end{figure}

\begin{figure}[ht]
\begin{center}
\epsfig{figure=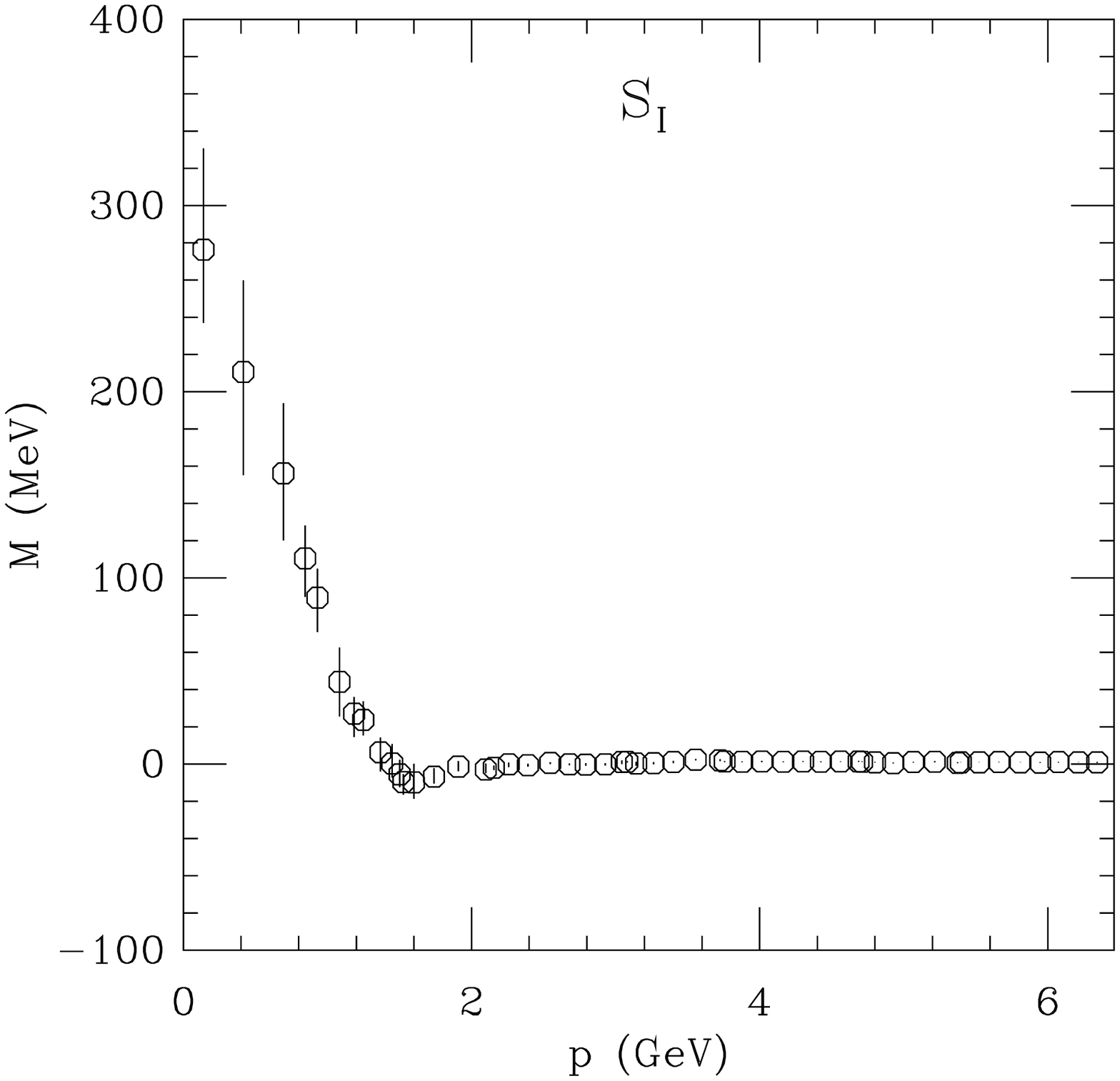,angle=0,width=6cm}
\end{center}
\caption{The tree-level corrected mass function from $S_I$ with
$\csw=\text{NP}$, with the bare mass extrapolated
to zero using a
quadratic fit.  The small dip at $p\sim1.6$ GeV is not statistically
significant and may be due to residual lattice artifacts.  The
non-zero values for $M(p)$ in the chiral limit are entirely due
to dynamical chiral symmetry breaking and provide a direct measure of
this effect.}
\label{fig:Mall-chiral}
\end{figure}

In the continuum, the quark propagator has the following general form,
\begin{equation}
S(p) = \frac{1}{i\pslash A^c(p) + B^c(p)} \equiv
\frac{Z^c(p)}{i\pslash+M^c(p)}.
\end{equation}
We expect the lattice quark propagator to have a similar form, but
with $\kslash$ replacing $\pslash$:
\begin{equation}
S(p) = \frac{Z(p)}{i\kslash + M(p)}
\end{equation}
where $k$ is a new `lattice momentum', 
$k_\mu = \frac{1}{a}\sin(\hat p_\mu a)$.
We do not have sufficient space here to describe the hybrid
tree-level correction that was used for the quark propagator
results presented here, but a detailed description has recently
been given.\cite{quark_prop}  We again use the cylinder cut
to further reduce hypercubic discretization artefacts.  
As for the gluon propagator the results for $Z(p)$ are for the
bare quantity only and contain an overall renormalization constant
$Z_2(\mu,a)$.  In Fig.~\ref{fig:z_np_compare}
the vertical scales have been adjusted so that the two sets of results
are renormalized and hence coincide at 2.1~GeV.
In this figure we see the charactreistic dip in the infrared for $Z(p)$,
which occurs also in model Dyson-Schwinger equation
studies\cite{DSE_review} of dynamical
chiral symmetry breaking.  This dip has essentially disappeared by
around 2~GeV.  The improved action correspondning to $S_R$ is the
one we prefer and it gives the more expected ultraviolet behavior
of the renormalized $Z(p)$, i.e., it tends towards a constant.  
In Fig.~\ref{fig:m_np_phys} we see the characteristic behavior
of the quark mass function familiar from quark model
studies\cite{DSE_review} and that the transition to the perturbative
regime is occuring at approximately 2 GeV.  Note that there is no
renormalization of the quark mass function.  At large momenta
the mass function should become the running quark mass of perturbative
QCD.  Finally in Fig.~\ref{fig:Mall-chiral} we present a simple quadratic
extrapolation to the chiral limit for the available $S_I$ data.
The slight dip is not statistically significant and is almost
certainly a residual lattice artefact.  The infrared mass (i.e., at
$p=0$) in the chiral limit is approximately $330\pm30$~MeV, which is
characteristic of the constituent quark mass scale.

\section{Summary and Conclusions}

The gluon propagator has been calculated on fine unimproved lattices
and on a variety of improved lattices with an
\oa{2} improved action in \oa{2} improved Landau gauge.  The infrared
behavior of this propagator strongly suggests the the Landau gauge
gluon propagator is infrared finite.  
We have ruled out the $1/q^{4}$ behavior popular in some Dyson-Schwinger
quark model studies\cite{DSE_review}
and indeed any infrared singularity appears to be very unlikely.
The possible effects of lattice Gribov copies remains a very interesting
question and we are currently carrying out similar studies across
a variety of lattices in Laplacian gauge, which is a Landau-like
smooth gauge fixing, but is free of Gribov copies.

We have used two different definitions of the ${\cal O}(a)$ improved
quark propagator, corresponding to the quark propagators
denoted $S_I$ and $S_R$.
We make use of asymptotic freedom to factor out the
tree level behaviour, replacing it with the `continuum' tree level
behaviour $Z(p)=1, M(p)=m$.  This
tree-level correction dramatically improves the data.  
We find that $M(0)$ approaches a value of $300\pm30$ MeV in the
chiral limit, which is very much in keeping with the concept
of a ``constituent quark mass'' and agrees with the infrared values
of the quark mass commonly used in model studies.\cite{DSE_review}
We also find a significant dip in the value for $Z(p)$ at low
momenta.  This is again entirely consistent with what
is found in model studies of dynamical chiral symmetry breaking
\cite{DSE_review}.  An examination of
Figs.~\ref{fig:Comp1i_6}, \ref{fig:pert_vs_latt}, \ref{fig:z_np_compare},
\ref{fig:m_np_phys}, and \ref{fig:Mall-chiral} provide a clear indication
that perturbative QCD behavior is not becoming dominant in the gluon
and quark propagators until we reach momenta of order
$Q^2\simeq 4$~GeV$^2$.



\end{document}